\documentclass[preprint,prd,aps,tighten,nofootinbib,amssymb]{revtex4}

\usepackage{graphicx}
\usepackage{epsfig,amssymb,amsmath}

\pdfoutput=1
\usepackage{subfigure}
\usepackage[colorlinks=true,linkcolor=blue,citecolor=blue]{hyperref}
\usepackage{float}

\makeatletter
\def\Hline{%
\noalign{\ifnum0=`}\fi\hrule \@height 2pt \futurelet
\reserved@a\@xhline}
\makeatother

\newcommand{\beq}{\begin{equation}}
\newcommand{\eeq}{\end{equation}}
\newcommand{\bea}{\begin{eqnarray}}
\newcommand{\eea}{\end{eqnarray}}
\newcommand{\bear}{\begin{array}}
\newcommand {\eear}{\end{array}}
\newcommand{\bef}{\begin{figure}}
\newcommand {\eef}{\end{figure}}
\newcommand{\bec}{\begin{center}}
\newcommand {\eec}{\end{center}}

\def\lrfp#1#2#3{ \left(\frac{#1}{#2} \right)^{#3}}

\begin{document}
\draft
\tighten
\preprint{TU-993,~IPMU15-0059}
\title{\large \bf
Axion domain wall baryogenesis
}
\author{
Ryuji Daido\,$^{a}$\footnote{email:daido@tuhep.phys.tohoku.ac.jp},
Naoya Kitajima\,$^{a}$\footnote{email:kitajima@tuhep.phys.tohoku.ac.jp},
Fuminobu  Takahashi\,$^{a,b}$\footnote{email: fumi@tuhep.phys.tohoku.ac.jp}
}
\affiliation{
$^a$ Department of Physics, Tohoku University, Sendai 980-8578, Japan\\
$^b$ Kavli IPMU, TODIAS, University of Tokyo, Kashiwa 277-8583, Japan
}

\vspace{2cm}

\begin{abstract}
We propose a new scenario of baryogenesis, in which annihilation 
of axion domain walls generates a sizable baryon asymmetry. 
Successful baryogenesis is possible for a wide range of the axion mass and decay constant,
$m  \simeq 10^8 -10^{13}$\,GeV and $f \simeq 10^{13} - 10^{16}$\,GeV.
Baryonic isocurvature perturbations are significantly suppressed in our model, in contrast to various 
spontaneous baryogenesis scenarios in the slow-roll regime. In particular, 
the axion domain wall baryogenesis  is  consistent with high-scale inflation which generates
a large tensor-to-scalar ratio within the reach of future CMB B-mode experiments.
We also discuss the gravitational waves produced by the domain wall annihilation and its
implications for the future gravitational wave experiments. 
\end{abstract}

\maketitle

\section{Introduction}
\label{sec:intro}
Axions may be ubiquitous in nature. Indeed, there appear many axions through compactification 
of the extra dimensions in the string theory~\cite{Svrcek:2006yi,Arvanitaki:2009fg}. Some of them may 
remain relatively light and play an important role in cosmology such as inflation, dark matter and dark energy.
In this paper we shall present a new scenario of baryogenesis, in which  axions play a key role.

The axion exhibits a shift symmetry,
\begin{align}
a \;\rightarrow\;  a + C,
\label{shiftsym}
\end{align}
where $C$ is a real transformation parameter. While the shift symmetry 
keeps  the axion potential flat at the perturbative level, non-perturbative
effects break the symmetry to a remnant discrete one.

Let us suppose that one of the non-perturbative effects gives the dominant contribution to the axion 
potential, which is expressed as
\beq
V(a)\;\simeq\; m^2 f^2 \left(1-\cos \left(\frac{a}{f}\right) \right),
\eeq
where $m$ is the axion mass and $f$ is the decay constant.
Then, the axion potential has a series of $N$ (approximately) degenerate vacua, where
the precise value of $N$ depends on the details of the UV theory.\footnote{The fact that the axion 
potential can have multiple approximately degenerate vacua has been exploited in context of
dark energy~\cite{Banks:1991mb} as well as inflation~\cite{Czerny:2014wza,Kallosh:2014vja,Higaki:2014pja}.}
If the axion is lighter than the Hubble parameter during inflation, it acquires quantum fluctuations which
extend beyond the Hubble horizon. For sufficiently large quantum fluctuations, some of the $N$ vacua 
might be populated, which results in domain wall formation after inflation. The domain walls are cosmologically problematic,
and so, they must annihilate before dominating the Universe.
 This is possible if the degeneracy between different vacua is lifted by other non-perturbative 
effects~\cite{Vilenkin:1981zs,Gelmini:1988sf,Coulson:1995nv,Larsson:1996sp}. The domain wall annihilation and the emitted gravitational waves have been extensively studied in the 
literature~\cite{Gleiser:1998na,Hiramatsu:2010yz,Kawasaki:2011vv,Hiramatsu:2013qaa}.

In this paper we point out that the annihilation of domain walls also induces the baryon asymmetry
of the Universe. Suppose that the axion is derivatively coupled to the standard model (SM) quarks 
and/or leptons,
\beq
{\cal L} = \frac{\partial_\mu a}{f} j^\mu =  \sum_i c_i \frac{\partial_\mu a}{f} \bar \psi_i \gamma^\mu \psi_i,
\label{eq:derivative_coupling}
\eeq
where $c_i$ is a coupling constant.
 The time derivative of the axion plays a role of the effective 
chemical potential, which spontaneously breaks the CPT symmetry.\footnote{
See Ref.~\cite{Bolokhov:2006wu} for leptogenesis using explicit (non-dynamical) CPT-breaking interactions.
} This enables the generation of the
baryon or lepton asymmetry in thermal plasma if the baryon or lepton number is broken,
and this is the so-called spontaneous baryogenesis scenario~\cite{Cohen:1987vi,Dine:1990fj,Cohen:1991iu}. 
 The current to which the axion is coupled 
does not have to coincide exactly with the baryon or lepton current; for instance,
it could be a U(1) hypercharge current~\cite{Cohen:1991iu} or a Peccei-Quinn 
current~\cite{Chiba:2003vp}. Such derivative couplings to the baryon and lepton currents
can also be induced if the axion has an anomalous coupling to the SU(2) 
gauge fields~\cite{Dine:1990fj}. In this case the chemical potential
is induced by sphalerons \cite{Arnold:1987mh,Ringwald:1987ej}, because a non-zero time derivative of the axion generates energy difference
between the states with different winding number and $B+L$ number.
Therefore, the chemical potential is expected to be suppressed at $T \gtrsim 10^{12}$\,GeV where sphalerons decouple from the cosmic expansion. Note that there is no such suppression of the effective chemical potential 
if one starts with the derivative couplings with baryon and/or lepton current (more precisely, $B-L$ current), as we shall do below.
We shall see that, if the axion has such derivative couplings,
 a sizable baryon asymmetry can be generated when the axion domain walls annihilate.

Before going into details,  let us give a rough sketch of our scenario.
For simplicity, we assume that only two vacua, $a_1$ and $a_2$ with
$a_1 < a_2$, are populated during inflation, leading to formation of domain walls separating the two vacua.
Generalization to the case of multiple vacua is straightforward.  After formation,
 domain walls randomly move around at relativistic speed,  collide and annihilate continuously, so that
the domain wall network show the dynamical scaling behavior~\cite{Press:1989yh,Hindmarsh:1996xv,Garagounis:2002kt,Leite:2011sc}. 
Every time a domain wall goes through some point in space,
the field value of the axion 
changes either from $a_1$ to $a_2$ or from $a_2$ to $a_1$. 
 Such transition induces a temporal and local chemical potential for baryons or leptons.
No net baryon asymmetry is generated by the domain wall dynamics in the scaling regime, however, 
because both transitions occur with an equal probability and there is no preference of baryons over anti-baryons. 
The asymmetry between the two vacua becomes important when the domain walls annihilate because of the
energy bias. Suppose that one of the vacua is energetically preferred, e.g., $V(a_1) < V(a_2)$.
When domain walls annihilate, the value of $a$ then decreases from $a_2$ to
$a_1$ in a region of the false vacuum, which gives a preference to baryons over anti-baryons for a certain
choice of the couplings.  Thus, the axion domain wall annihilation can generate the baryon asymmetry of the Universe. 

Our scenario has several advantages. First, it is known that  the spontaneous baryogenesis in the slow-roll regime 
generically leads to  baryonic isocurvature perturbations~\cite{Turner:1988sq}, which makes the scenario incompatible 
with high-scale inflation.\footnote{It is possible to give the axion a mass of order the Hubble parameter in the 
spontaneous baryogenesis using a flat direction~\cite{Chiba:2003vp}, thus avoiding the isocurvature constraint. 
Also, no isocurvature perturbation is induced in the gravitational baryogenesis \cite{Davoudiasl:2004gf}.
} 
In our scenario, however,  the baryonic isocurvature perturbations can be 
significantly suppressed, because of the scaling property of the domain wall network.
In particular, our scenario is consistent with large-field inflation, and therefore,  the required high reheating temperature can be realized more easily. 
Secondly, the axion field value is kept large inside domain walls, which enables a large effective chemical
potential even when the axion mass $m$ becomes larger than the Hubble parameter. Without domain walls,
the spontaneous baryogenesis  would become inefficient when the axion starts to oscillate about the 
minimum~\cite{Dolgov:1996qq}. Therefore, the axion domain wall 
baryogenesis scenario  works for a wide range of the axion mass
and the inflation scale.

Lastly let us comment on differences of our scenario from other works. In the thick-wall regime of the
electroweak baryogenesis, the passage of an expanding bubble wall generates a non-zero chemical potential,
which leaves net baryon asymmetry in thermal plasma based on the spontaneous baryogenesis~\cite{Dine:1990fj,Cohen:1991iu} (see also Ref.~\cite{Cheung:2012im}).
The bubble walls play a similar role to that of domain walls in our scenario. The difference is that
the electroweak spontaneous baryogenesis relies on the first order phase transition of two (or more) Higgs fields, and
 the sphaleron process is exponentially suppressed in the symmetry breaking vacuum. As a result, the estimate
 of the final baryon asymmetry requires a precise  determination of the critical field value as well as detailed analysis 
 of the diffusion process during the phase transition~\cite{Cohen:1994ss}. 
In our scenario, on the other hand, the baryon (or lepton) number violation is operative equally in the two minima.
Also it relies
on the domain wall dynamics of a single axion field, whose behavior is well studied with numerical simulations. This makes our scenario
 relatively simple and robust. 
Recently, the authors of Ref.~\cite{Kusenko:2014uta} proposed a  scenario where the axion has only anomalous coupling 
to SU(2)$_L$ gauge fields. They studied a spatially homogeneous axion field 
in the slow-roll regime, and explored the parameter space of the axion mass and decay constant preferred by the string axions.  The parameter ranges have an overlap with our scenario. One difference is that
 we start with derivative couplings of the axion with baryon and/or lepton currents.
Another is that our scenario relies
on the domain wall dynamics, while Ref.~\cite{Kusenko:2014uta} focused on the homogeneous axion field.

The rest of this paper is organized as follows. In Sec.~\ref{sec:adw}, we briefly review the evolution of axion domain walls. We estimate
the baryon asymmetry induced by the domain wall annihilation in Sec.~\ref{sec:bg}. The last section is devoted to discussion and conclusions.

\section{Axion domain walls}
\label{sec:adw}

Let us consider an axion whose potential is given by
\beq
V(a) \;=\; m^2 f^2 \left( 1-\cos\left(\frac{a}{f}\right) \right),
\label{Va}
\eeq
where $m$ and $f$ are the mass and the decay constant of the axion $a$.
We assume that two adjacent minima, $a_1 = 0$ and $a_2 = 2 \pi f$,  are populated with more or less equal probability during inflation, and that domain walls separating the two minima are formed after inflation. 
This is the case if the quantum fluctuations of the axion, $\delta a \sim H_\mathrm{inf}/2\pi$, is comparable
to the decay constant, or if the initial position of the axion is sufficiently 
close to the local maximum, $a_\mathrm{max} = \pi f$.
Our scenario can be straightforwardly applied to the case in which more than two minima
are populated. 

The domain wall solution in a flat spacetime is given by
\beq
\label{dwsol}
	a_\mathrm{dw}(t,\vec{x}) = 4f \tan^{-1} \exp \big[ m \gamma (x-v t) \big],
\eeq
where $x$ is the spatial coordinate perpendicular to the domain wall, 
$v$ is the domain wall velocity and $\gamma$ is the relativistic factor defined by $\gamma = 1/\sqrt{1-v^2}$.
The above solution is valid  if the thickness of the domain wall $\sim m^{-1}$
is much smaller than the Hubble horizon, i.e., $m \gg H$, where $H$ is the Hubble parameter. 
The energy density of the domain wall is characterized by the tension $\sigma$, 
\beq
\sigma = 8 m f^2,
\eeq
for the potential (\ref{Va}).

The domain walls are formed when $H \simeq m$.
 According to the numerical and analytic calculations~\cite{Press:1989yh,Hindmarsh:1996xv,Garagounis:2002kt,Leite:2011sc}, 
within a  few Hubble time after the formation, the domain walls quickly follow
the scaling law, i.e., 
\beq
\rho_\mathrm{dw} \sim \sigma H,
\eeq
where there are only one or a few domain walls in each Hubble horizon.
The domain walls must annihilate and disappear before they start to dominate the Universe, since otherwise
the Universe would be too inhomogeneous. We assume that there is another shift-symmetry
breaking term which generates a bias between the two minima, $\epsilon \equiv V(a_2)- V(a_1)$.
Then domain walls annihilate rapidly when the energy density of domain walls becomes
comparable to the energy bias \cite{Gelmini:1988sf,Coulson:1995nv,Larsson:1996sp},
\beq
\rho_\mathrm{dw} \sim \epsilon.
\eeq
Marginally relativistic axion particles with a typical momentum, $k \sim m$, are
copiously produced through the axion domain wall annihilation. Those axion particles
soon become non-relativistic due to the cosmic expansion~\cite{Hiramatsu:2010yz,Kawasaki:2011vv,Hiramatsu:2013qaa}.
In addition,  axion coherent oscillations are produced at the domain wall formation, and
we shall discuss their cosmological impact later in this paper. 

The axion particles
eventually decay into SM particles through their couplings with the SM sector.
In general, the axion can have derivative couplings to fermions like (\ref{eq:derivative_coupling}),
which are allowed by the shift symmetry (\ref{shiftsym}).
Specifically we focus on the case in which
 the axion has derivative couplings only to the SM left-handed lepton currents,\footnote{
In a supersymmetric theory, this type of coupling arises from the K\"ahler potential
$K = \frac{1}{f}(\mathcal{A}+\mathcal{A}^\dagger) L^\dagger L$,
where $\mathcal{A} $ and $L$ are respectively the axion and the lepton supermultiplet, and the lowest component of $\mathcal{A}$
is given by the saxion and axion as $\mathcal{A}=s+i a$.
}
\beq
	\mathcal{L} \ni \frac{\partial_\mu a}{f} \sum_{i=e,\mu,\tau} \bar{L}_i \gamma^\mu L_i \equiv \frac{\partial_\mu a}{f} j^\mu.
	\label{eq:dela_jL}
\eeq
Our results remain practically unchanged even if one adds additional derivative couplings
to other SM fermions.
If the axion is coupled to the SM sector only through the above interaction (\ref{eq:dela_jL}), 
it mainly decays into a pair of SU(2)$_L$ gauge bosons and hypercharge gauge bosons through its 
anomalous couplings~\cite{Cohen:1987vi}. The decay width into a pair of gauge bosons is approximately 
given by
\beq
	\Gamma_a \simeq \bigg( \frac{3\alpha_2^2}{256 \pi^3} + \frac{\alpha'^2}{1024 \pi^3} \bigg) \frac{N_f^2 m^3}{f^2},
\eeq
where $\alpha_2$ and $\alpha'$ are respectively SU(2)$_L$ and U(1)$_Y$ gauge coupling constants and $N_f $ is the number of of generation, and we will set $N_f = 3$ in the following.
Approximating that this is the main decay channel, the axion decay temperature is
\beq
	T_{a} \simeq 3 \times 10^7~\mathrm{GeV} \bigg(\frac{m}{10^{11}\,\mathrm{GeV}} \bigg)^{3/2} \bigg( \frac{10^{15}\,\mathrm{GeV}}{f} \bigg),
\eeq
where we have defined the decay temperature by $3H(T_a) = \Gamma_a$.
If those axion particles dominate the Universe before the decay, there will be an extra entropy production by the axion decay,
which dilutes  pre-existing baryon asymmetry by some amount.
As we shall see, the entropy dilution becomes important for a large decay constant and a small axion mass.

\section{Baryogenesis by domain wall annihilation}
\label{sec:bg}

\subsection{Analytical estimate of the asymmetry}
Now let us discuss  baryogenesis by the axion domain walls under the existence of the 
derivative coupling to the lepton current given by (\ref{eq:dela_jL}).\footnote{
Instead, one may use an anomalous coupling of the axion to the SU(2) gauge fields, in which the baryogenesis works similarly as long as the sphalerons are in equilibrium at the domain wall annihilation. (See the discussion in Sec.~\ref{sec:intro}).}
As previously noted, if $\dot{a}$ is non-vanishing,  the derivative couplings behave like an effective chemical potential,
\beq
\frac{\partial_\mu a}{f} j^\mu = \mu_\mathrm{eff} j^0 + \dots,
\eeq
where  $\mu_\mathrm{eff} = \dot{a}/f$ is the effective chemical potential for the lepton  number ($L$).

The axion domain walls can generate the effective chemical potential because of the large spatial gradient of the axion field inside the wall.
Since domain walls are moving at nearly the speed of light, the time derivative of the axion field at some fixed spatial point  becomes large while domain walls are passing through. 
The effect of the gradient term is negligible if the domain wall is sufficiently thick compared to the
diffusion length.

If the $L$-number violating operator is in equilibrium, and if the chemical potential is spatially homogeneous, 
the difference of number densities between lepton and anti-leptons would be produced as
 $n_\ell^\mathrm{eq} - n_{\bar{\ell}}^\mathrm{eq} \simeq 2 \mu_\mathrm{eff} T^2$ 
 for $\mu_\mathrm{eff} \ll T$, where we have taken into account the spin degrees of freedom 
 and the number of generation.
 It depends on the rate of the $L$-number violating process as well as the domain wall dynamics whether the lepton asymmetry reaches
 the equilibrium value in the expanding Universe. One needs to solve
the Boltzmann equation for the lepton asymmetry, $n_L = n_\ell - n_{\bar{\ell}}$, 
\beq
	\dot{n_L} + 3H n_L = -\Gamma (n_L - n_L^\mathrm{eq}),
	\label{eq:BoltzmannEq}
\eeq
where $\Gamma$ is the interaction rate for the $L$-violating processes. Note here that the chemical potential in $n_L^\mathrm{eq}$
depends on the position and velocity of domain walls.

As the $L$-number violating operator, we consider $\Delta L = 2$ scattering processes, $\ell \ell \leftrightarrow H H$, 
$\ell H \leftrightarrow \bar\ell \bar{H}$,  which are mediated by  heavy right-handed Majorana neutrinos
in the seesaw mechanism~\cite{Minkowski:1977sc,Yanagida:1979as,Ramond1979,Glashow:1979nm}.
Here and in what follows we assume that the right-handed neutrinos are so heavy that they can be
integrated out in our analysis.
The interaction rate for the $\Delta L = 2$ processes is roughly given by~\cite{Buchmuller:2000nq}
\beq
	\Gamma \sim \frac{T^3}{\pi^3} \frac{\sum m_i^2}{v_\mathrm{EW}^4},
	\label{DeltaL2}
\eeq
where $v_{\mathrm EW} = 174$~GeV and $m_i$ with $i=1,2,3$ denotes the mass of three active neutrinos.
The decoupling temperature of the $L$-violating process in the radiation dominated Universe is 
\beq
\label{decT}
T_\mathrm{dec} \sim  3 \times 10^{13}\,\mathrm{GeV},
\eeq
where we have assumed the normal ordering for the neutrino mass differences and used the experimental value, $\sum m_i^2 \simeq \Delta m^2_\mathrm{atm} \simeq 2.4 \times 10^{-3}~\mathrm{eV}^2$.
For the reheating temperature $T_R$ lower than $T_\mathrm{dec}$, the $L$-violating process remains 
decoupled from the cosmic expansion. As we shall see below, even in this case, a non-zero lepton asymmetry is induced
by the domain wall annihilation.

 Let us first consider an ideal situation where a domain wall passes through
the origin $\vec{x} = 0$ at $t=t_{DW}$. Using Eq.~(\ref{dwsol}), the effective chemical potential at the origin evolves with time as
\beq
	\mu_\mathrm{eff} = -\frac{2 m \gamma v}{\cosh[m\gamma v(t-t_\mathrm{DW})]}.
	\label{chem}
\eeq 
It takes roughly $\Delta t \sim (m\gamma v)^{-1}$ for the domain wall to pass through the origin, and
so,  the induced lepton asymmetry by passage of the domain wall is estimated as
\beq
	n_L \simeq  \Gamma n_L^\mathrm{eq} \Delta t \sim \Gamma T^2.
	\label{nLsingle}
\eeq
Note that the lepton asymmetry becomes independent of the velocity of the domain walls.
As the domain wall passes through, a similar amount of the lepton number density will be induced 
inside the Hubble horizon.

In the scaling regime, domain walls randomly move around inside the Hubble horizon so as to 
collide and annihilate continuously. In particular,  since there is no preference for either of the vacua,
the effective chemical potential can be positive or negative with  equal probability.  Therefore there will be no net
lepton asymmetry left, even though some amount of the lepton asymmetry with either positive or negative sign
is induced each time a domain wall passes through.  
Such lepton asymmetry has  fluctuations of order unity inside the Hubble horizon, but it has no sizable 
fluctuations at superhorizon scales, because of the scaling property of the domain-wall network.

A non-zero net lepton asymmetry is induced when domain walls annihilate and disappear owing to the energy 
bias. This is because one of the two vacua is energetically preferred, inducing an effective chemical potential with a fixed sign in 
the false vacuum which occupies about half of the space. Again, the scaling property of the domain wall network ensures  that there is no isocurvature perturbations at super-horizon scales. 

The final lepton asymmetry is generated within about one or a few Hubble time before the domain wall annihilation.
In particular, the maximal possible value of the lepton asymmetry is obtained when the domain wall annihilation takes place at the decoupling of the $L$-violating processes. The reason is as follows. If the domain wall annihilation takes place before the decoupling of the $L$-violating processes, the lepton asymmetry induced by the domain wall annihilation will be washed out. On the other hand, if the domain wall annihilation occurs after the decoupling, the induced asymmetry tends to be suppressed because the $L$-violating process is inefficient. 
The maximum asymmetry is therefore
\beq
	\frac{n_L}{s}\bigg|_\mathrm{max} 
	\simeq -\frac{45}{\pi^2 g_{*s}} \frac{\Gamma}{T}\bigg|_\mathrm{dec} 
	\sim -  10^{-6},
	\label{eq:nLs_max}
\eeq
where $s$ and $g_{*s}$ are respectively the entropy density and the relativistic degrees of freedom.
We have substituted $g_{*s} = 106.75$ and the decoupling temperature (\ref{decT}) in the second equality,
assuming the radiation-dominated Universe. The negative sign is inserted in the second equality 
to obtain positive baryon asymmetry through sphalerons. 

If the reheating temperature $T_R$ is lower than the decoupling temperature $T_\mathrm{dec}$,
 the interaction rate for the $L$-violating processes never exceeds the expansion rate of the Universe.  One can see this by noting that
$\Gamma/H$ reaches the maximal value (smaller than unity) at the reheating as long as
the temperature of the dilution plasma obeys $T \sim (H T_R^2 M_P)^{1/4}$ before the reheating.
Hence the maximal asymmetry in this case is obtained if the domain wall annihilation occurs at the reheating,
and it is roughly given by 
\beq
\left.\frac{n_L}{s}\right|_R \simeq \left.\frac{n_L}{s}\right|_\mathrm{max} \lrfp{T_R}{T_\mathrm{dec}}{2}.
\eeq
We shall see later in this section that the maximal asymmetry is indeed generated if the domain wall
annihilation takes place at $T = \mathrm{min}[T_{\rm dec}, T_R]$.

\subsection{Necessary conditions for successful baryogenesis}
Here let us discuss some necessary conditions for the successful domain wall baryogenesis.
First, the domain wall dynamics should have negligible back reaction from the generated lepton asymmetry in the plasma.
As the domain walls move in the plasma, some amount of the lepton asymmetry is induced because of the effective
chemical potential (\ref{chem}). The interaction with the generated asymmetry induces a back reaction, which would
act as a frictional force on the domain wall dynamics. The back reaction is negligible, and the domain walls follow the scaling law if
\beq
\sigma H \gtrsim \mu_\mathrm{eff} n_L
\label{sH}
\eeq
 at the domain wall formation ($H_\mathrm{form} \sim m$), where $n_L$ is given by (\ref{nLsingle}).

Secondly, the domain wall must be sufficiently thick to justify our analysis where we have neglected 
dissipation of the asymmetry.
The thickness of the wall is roughly $m^{-1}$ and the typical mean free path of the particle in plasma 
is of order $T^{-1}$. Thus, the thick-wall condition is given by 
\beq
T_\mathrm{ann} > m,
\eeq
 where $T_\mathrm{ann}$ denotes the temperature at the domain wall annihilation.

Thirdly,  we have assumed that the domain wall annihilation 
 takes place well after the domain wall network start to follow
the scaling law. It takes  a few Hubble time after the formation to reach the scaling regime, and
therefore we conservatively  require 
\beq
H_\mathrm{form} \sim m > 10H_\mathrm{ann},
\label{m10H}
\eeq
where $H_\mathrm{form}$ and $H_\mathrm{ann}$ are the Hubble parameter at the domain wall
formation and annihilation, respectively.

Fourthly, we require that the decay constant is larger than the quantum fluctuations of the axion to ensure
the validity of analysis using the potential (\ref{Va}). Specifically, we impose a lower bound on $f$ as
\beq
f \gtrsim \frac{H_\mathrm{inf}}{2\pi},
\label{fH}
\eeq
where $H_\mathrm{inf}$ is the Hubble parameter during inflation. If this bound is not satisfied, the corresponding
U(1) symmetry may be restored, or the saxion field may be destabilized. 

Finally we assume that there is (effectively) only single path connecting the two vacua $a_1$ and $a_2$. 
Apparently this is not satisfied if a U(1) symmetry is explicitly broken down to $Z_2$. In this case there are
 two paths (clockwise and counter-clockwise) connecting  the two vacua. In other words,
 there appear two kinds of domain walls with the same number. This can be understood by
noting that the two types of the domain walls are attached to cosmic strings associated with the spontaneous break down of the U(1) symmetry.
If the tensions of the two type of domain walls are equal,  they would start to annihilate at the same time and sweep 
equal spatial volume with positive and negative chemical potential, resulting in no net baryon asymmetry.
On the other hand, if there is an explicit breaking of the $Z_2$ symmetry such that one type of domain walls has a larger tension than the other one,
the domain walls with a smaller tension would start to annihilate first by the energy bias between the two vacua and sweep a larger spatial region, 
producing a net baryon asymmetry. Therefore, our scenario works even if there are multiple paths connecting the two vacua (namely
if there are multiple types of domain walls), as long as one of the multiple paths is energetically favored.
 If there are multiple vacua, or if the symmetry is non-linearly realized, our scenario works by a similar argument. 
 
In the numerical calculations we impose the above conditions to ensure successful domain wall baryogenesis.
It turns out that all the conditions are easily satisfied for the parameters of our interest. 

\subsection{Numerical calculations}
The net lepton asymmetry is effectively induced by the domain wall annihilation, during which
 domain walls sweep typically  about a half of the space. To model the domain wall dynamics during 
 the annihilation, we approximated  the situation by a single domain wall passing through the origin, where
we numerically solve the Boltzmann equation (\ref{eq:BoltzmannEq}), combined with 
the evolution equations for the energy density of the inflaton ($\rho_I$) and  radiation ($\rho_r$),
\beq
	\dot{\rho_I} + 3H \rho_I = -\Gamma_I \rho_I,~~ \rho_r + 4H \rho_r = \Gamma_I \rho_I,
\eeq
where $\Gamma_I$ is the decay rate of the inflation, and we define the reheating temperature in our analysis 
by $3H(T_R) = \Gamma_I$. This approximation is valid because no net asymmetry is induced during the
scaling regime, and so, we can focus on the domain wall dynamics during the one or a few Hubble time
before the annihilation.

\begin{figure}[tp]
\centering
\subfigure[~$m = 10^{11}$ GeV]{
\includegraphics [width = 10cm, clip]{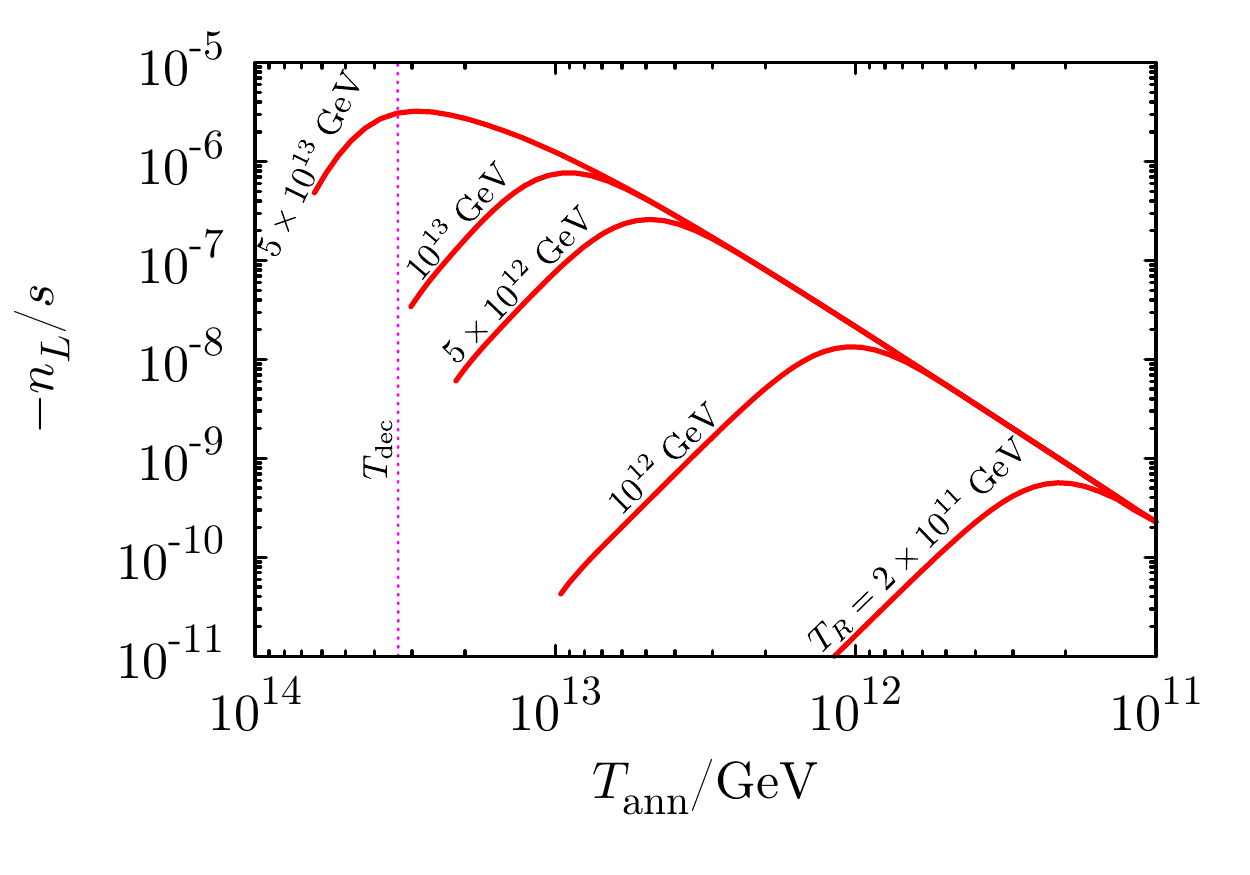}
\label{subfig:nLs-Tann_a}
}
\subfigure[~$m = 10^{12}$ GeV]{
\includegraphics [width = 10cm, clip]{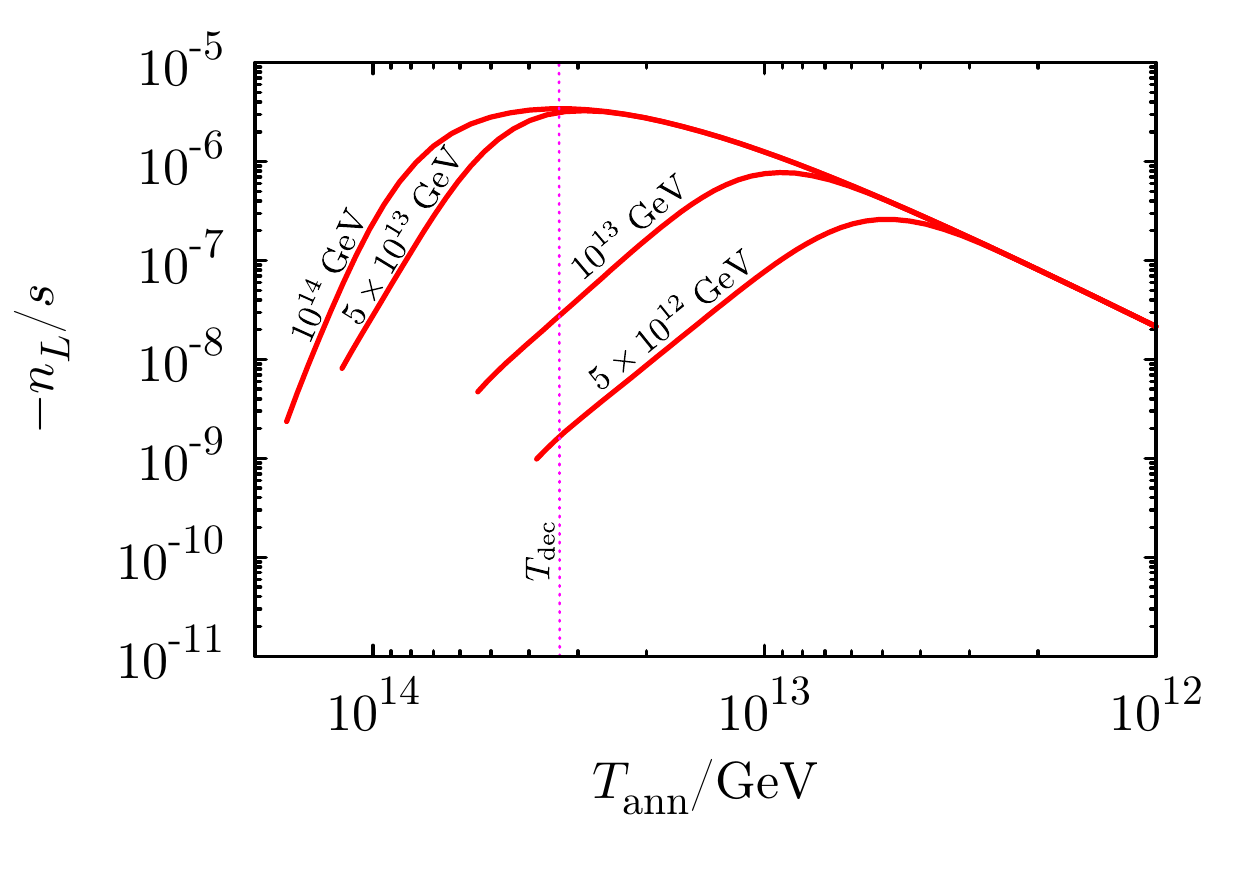}
\label{subfig:nLs-Tann_b}
}
\caption{
	The induced lepton asymmetry as a function of the domain wall annihilation temperature 
	for various values of $T_R$ and the axion mass  
	$m = 10^{11}$ GeV (top) and $10^{12}$ GeV (bottom).
	The vertical dotted (magenta) line represents the decoupling temperature of the $L$-number violating 
	processes in a radiation-dominated Universe. Note that the subsequent entropy dilution by the
	axion decay is not taken into account here.
	We have imposed the condition (\ref{m10H}), $m > 10H_\mathrm{ann}$, 
	which corresponds to the left end 
	point of each curve.
}
\label{fig:nLs-Tann}
\end{figure}

In Fig.~\ref{fig:nLs-Tann} we show the induced lepton asymmetry as a function of the domain wall annihilation temperature for various values of the reheating temperature. In the top and bottom panels, we have
set the axion mass to be $m=10^{11}$\,GeV and $10^{12}$\,GeV, respectively. 
Here we have not taken into account the
entropy production by the subsequent axion decay, which we shall return to in a moment. 
As expected, the maximal asymmetry is obtained when
$T_\mathrm{ann} \simeq \min (T_\mathrm{dec},T_R)$, in good agreement with the analytic estimate  (\ref{eq:nLs_max}).
In the bottom panel, one can see that the lepton asymmetry is highly suppressed in the case of
e.g.  $T_\mathrm{ann} > T_\mathrm{dec}$ and $T_R = 10^{14}$\,GeV. This is because the asymmetry
induced by the domain wall annihilation is subsequently washed out by the $L$-number violating processes in equilibrium. In general, we expect that the wash-out process is efficient when 
 $T_R > T_\mathrm{ann} > T_\mathrm{dec}$.

At the domain wall annihilation, marginally relativistic axions are copiously produced, and they may
come to dominate the Universe before they decay into gauge bosons.
Once the axion dominates the Universe, its subsequent decay produces a large entropy, diluting pre-existing
asymmetry.
Thus, the final baryon asymmetry is fixed after the axion decay, if there is entropy dilution. 
Taking into account the sphaleron process\footnote{
We have neglected  the sphaleron effects during the domain wall annihilation, for simplicity.
This approximation is valid for most of the parameters of our interest, because, as we shall see,
 successful baryogenesis requires $T_\mathrm{ann} \gtrsim 2 \times 10^{11}$\,GeV, while the sphalerons are
decoupled at $T\gtrsim 10^{12}$\,GeV.
Even if sphalerons are in equilibrium at the domain wall annihilation, 
the resultant baryon asymmetry changes only by a factor of ${\cal O}(1)$, and
our main results remain valid.
}, the resultant baryon asymmetry is estimated as
\beq
	\frac{n_B}{s} \simeq -\frac{28}{79} \times \frac{1}{2} \times \Delta \times \frac{n_L}{s} 
\eeq
where $\Delta$ is the dilution factor by the axion decay given by
\beq
	\Delta = \begin{cases} 
	\mathrm{min} \bigg( 1,~\frac{T_{a} H_\mathrm{ann} M_P^2}{T_R \sigma} \bigg) ~~&\text{(DW annihilation before reheating)} \\[4mm]
	\mathrm{min} \bigg( 1,~\frac{T_{a} s(T_\mathrm{ann})}{\sigma H_\mathrm{ann}} \bigg) ~~&\text{(DW annihilation after reheating)}
	\end{cases}.
\eeq
The numerical factor $1/2$ comes from the fact that the transition from the false vacuum to the true vacuum 
takes place  in about half of the whole space.

\begin{figure}[htp]
\centering
\subfigure[~$T_R = 2 \times 10^{11}$ GeV]{
\includegraphics [width = 9cm, clip]{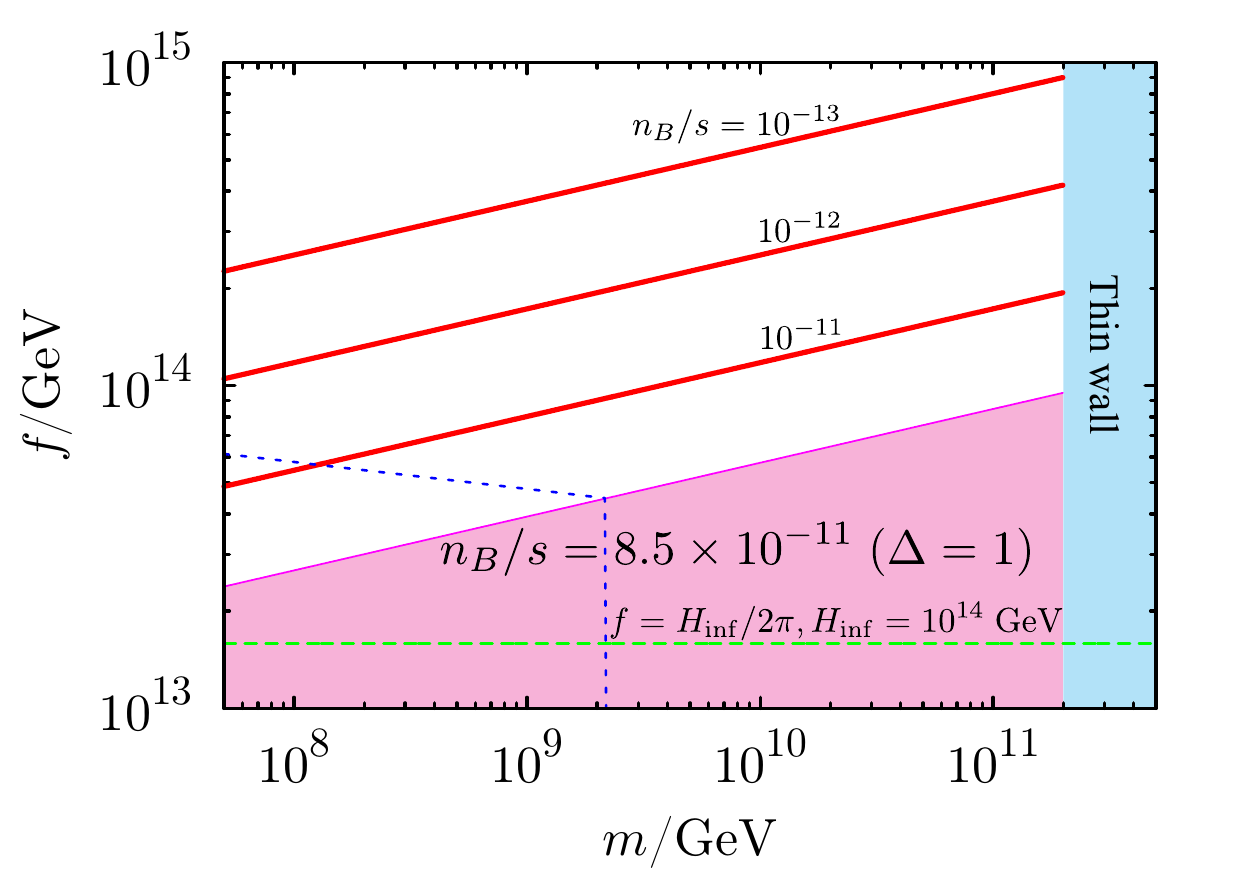}
\label{subfig:contour_a}
}
\subfigure[~$T_R = 10^{12}$ GeV]{
\includegraphics [width = 9cm, clip]{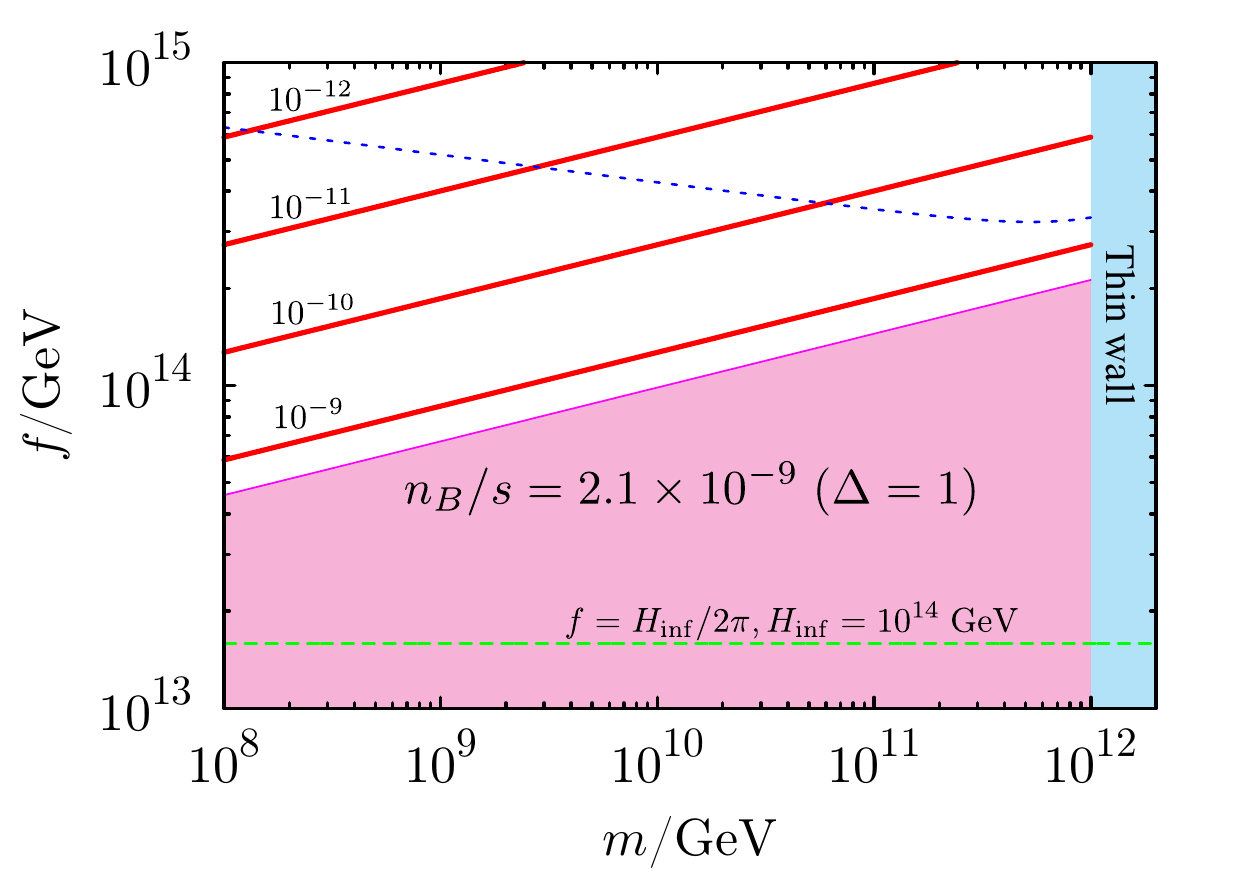}
\label{subfig:contour_b}
}
\caption{
Contours of the final (maximal) baryon number asymmetry in the 
$m$--$f$ plane for $T_R = 2 \times 10^{11}$ GeV (top panel) and $10^{12}$ GeV (bottom panel).
We assume $T_\mathrm{ann} = \mathrm{min}[T_R, T_\mathrm{dec}]$ so that
the baryon asymmetry  becomes maximal. 
The solid (red) lines correspond to the contours of $n_B/s = 10^{-13}$--$10^{-9}$ from top to bottom.
In the shaded (magenta) regions, there is no entropy dilution (i.e. $\Delta = 1$), and 
$n_B/s$ takes a constant value $n_B/s = 8.5\times 10^{-11}$ (top panel) and $2.1 \times 10^{-9}$ (bottom panel).
In the cyan-shaded region, the thick wall condition is violated.
Baryonic isocurvature perturbations and their non-Gaussianity will be too large below the dotted (blue) line, 
as long as one extrapolates the $L$-violating interactions to the domain wall formation. See the text for discussion on this issue.
The horizontal dashed (green) lines represent the lower bound on $f$, $f > \delta a \sim H_\mathrm{inf}/2\pi$ for $H_\mathrm{inf} = 10^{14}$ GeV.
}
\label{fig:contour_ab}
\end{figure}

\begin{figure}[h!]
\centering
\subfigure[~$T_R = 10^{13}$ GeV]{
\includegraphics [width = 9cm, clip]{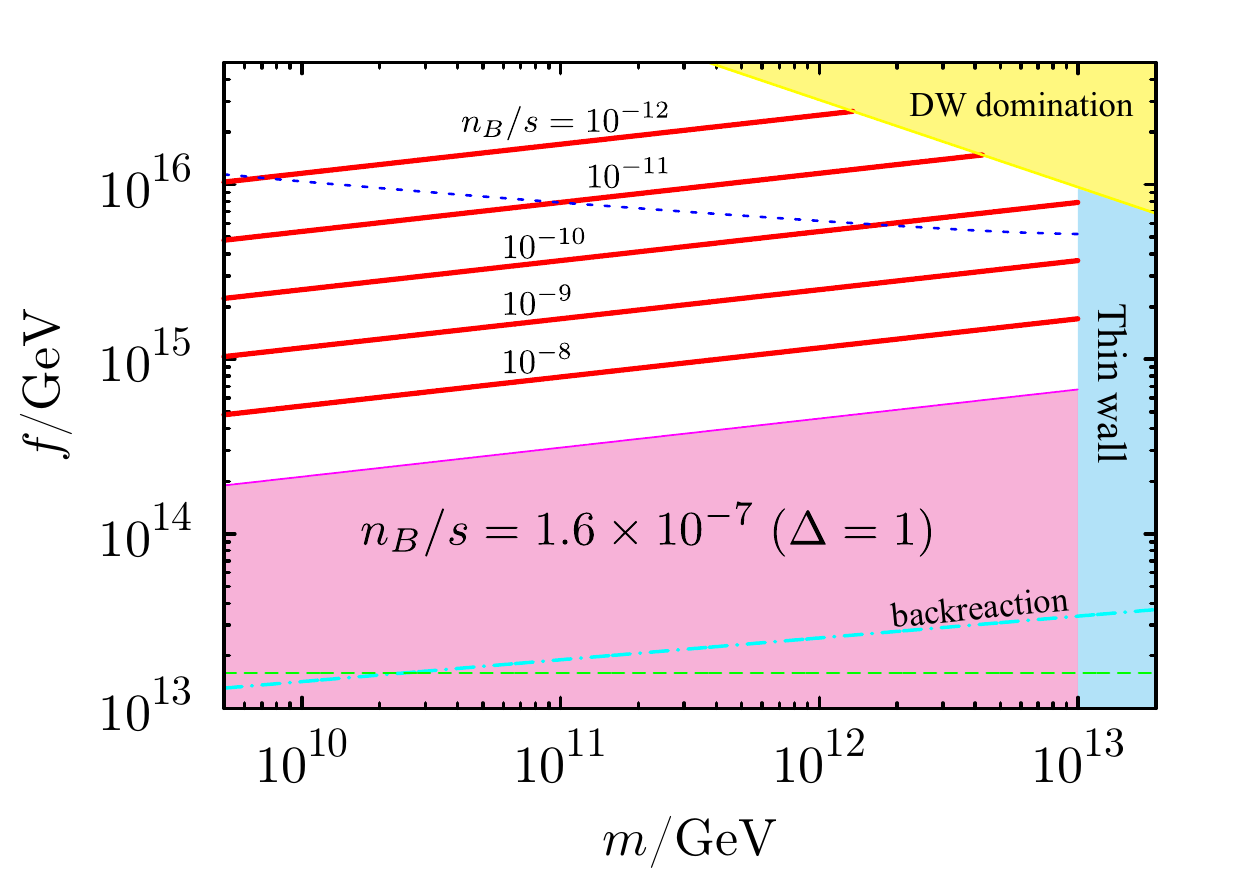}
\label{subfig:contour_c}
}
\subfigure[~$T_R = 10^{14}$ GeV]{
\includegraphics [width = 9cm, clip]{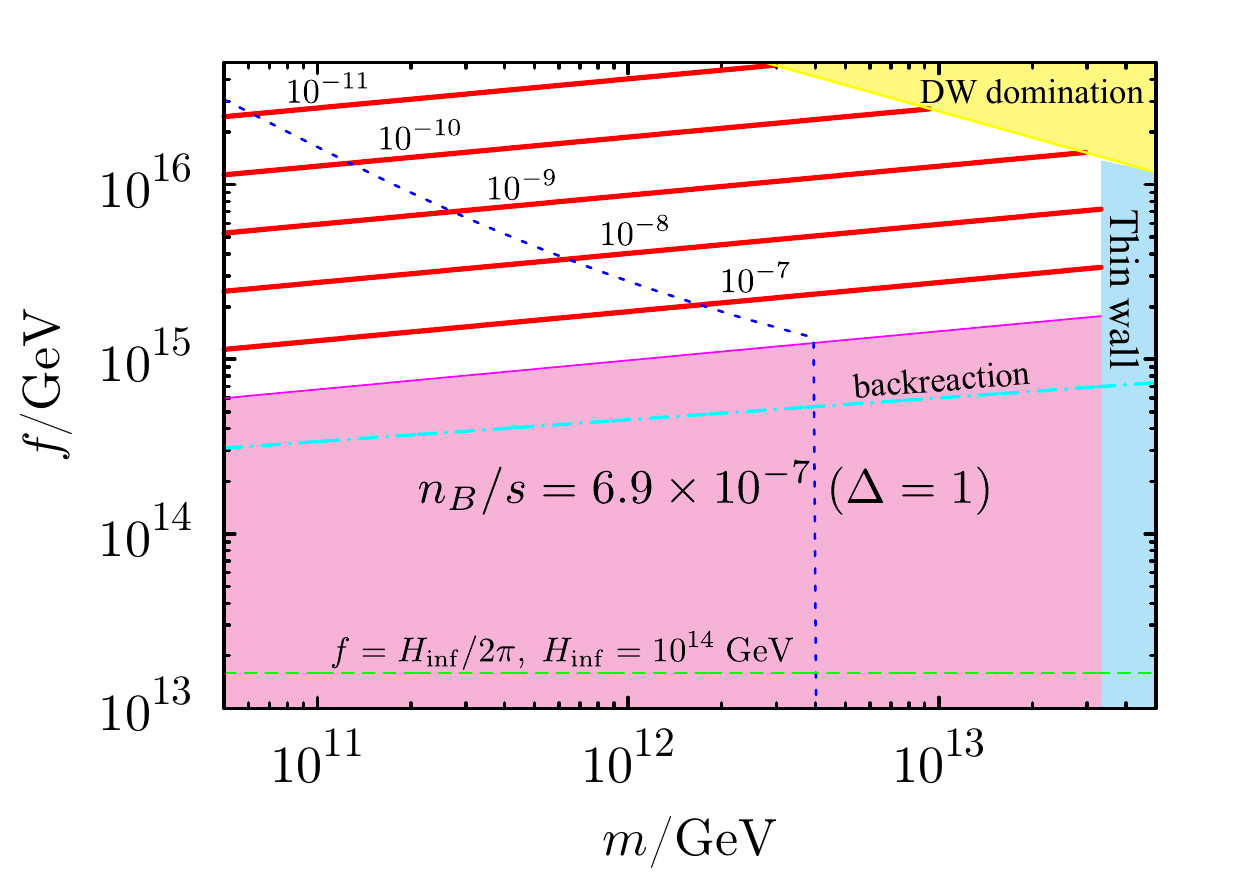}
\label{subfig:contour_d}
}
\caption{
	Same as Fig.~\ref{fig:contour_ab} but for $T_R = 10^{13}$ GeV (top panel), $10^{14}$ GeV (bottom panel).
	The solid red lines correspond to the contours of $n_B/s = 10^{-12}$--$10^{-7}$ from top to bottom and the magenta shaded regions correspond to the maximal value, $1.6\times 10^{-7}$ (top panel) and $6.9 \times 10^{-7}$ (bottom panel).
	The dash-dotted cyan line represents the lower bound from the back reaction and the yellow-shaded region is ruled out from domain wall domination.
}
\label{fig:contour_cd}
\end{figure}

In Figs.~\ref{fig:contour_ab} and \ref{fig:contour_cd} we show  the contours of  the final 
baryon asymmetry, $n_B/s$,  in the $m$--$f$ plane for various values of $T_R$. Here we have set $T_\mathrm{ann}
= \min (T_\mathrm{dec},T_R)$ so that the baryon asymmetry takes the largest possible value for a given
reheating temperature. 
The baryon asymmetry can be suppressed by either increasing or 
decreasing $T_\mathrm{ann}$ (see Fig.~\ref{subfig:nLs-Tann_a}).
One can see that a sufficient amount of baryon asymmetry, $n_B/s \gtrsim 10^{-10}$,
can be generated for $T_R \gtrsim 2 \times 10^{11}$ GeV. 
In the lower shaded (magenta) region, there is no entropy dilution, i.e., $\Delta \simeq 1$,
and so, $n_B/s$ takes a constant value. As $f$ becomes large, $n_B/s$ decreases owing to 
the entropy dilution factor $\Delta \ll 1$. This is because, as $f$ increases,  the energy density of the axion particles increases and 
 the lifetime of the axions becomes longer. The horizontal dashed (green) lines and dash-dotted (cyan) lines represent the lower bound 
 on the axion decay constant, $f \gtrsim H_\mathrm{inf}/2\pi$, for $H_\mathrm{inf} = 10^{14}$ GeV and $\sigma H > \mu_\mathrm{eff} n_L |_{H=m}$, respectively (cf. (\ref{sH}) and (\ref{fH})).
The yellow-shaded region in upper right corner in Fig.~\ref{fig:contour_cd} is ruled out from the domain wall domination at annihilation.
Below the dotted (blue) line, baryonic isocurvature perturbations and their non-Gaussianity would exceed the observational bound, 
 if the $L$-number violating rate (\ref{DeltaL2}) is valid at the domain wall formation. In other words, in the region slightly below 
 the dotted (blue) line, baryonic isocurvature perturbations and their non-Gaussianity
may be found in the near future observations. 
We will discuss this issue in the next subsection.

\subsection{Baryonic isocurvature perturbations}
Here we discuss  baryonic isocurvature perturbations in our scenario. Here we do not distinguish lepton asymmetry 
and baryon asymmetry, as we are concerned with the final baryon asymmetry at the CMB epoch.
First, let us consider baryon asymmetry generated by the domain wall annihilation,
$Y_\mathrm{DW,ann} \equiv n_{B}/s|_\mathrm{DW,ann}$. As domain walls are spatially localized objects, 
$Y_\mathrm{DW,ann}$ has initially large spatial fluctuations of order unity at subhorizon scales.
Such small-scale fluctuations asymptote to zero in the course of evolution, because
of diffusion processes of quarks and leptons. At super-horizon scales (e.g. the CMB scales),
on the other hand, $Y_\mathrm{DW,ann}$ has no isocurvature fluctuations because of the scaling
property of the domain wall network. This results stand in sharp contrast to the usual
spontaneous baryogenesis in the slow-roll regime~\cite{Turner:1988sq}.

Secondly, we turn to baryon asymmetry generated right after the domain wall
formation. We have assumed that the axion acquires sufficiently large quantum fluctuations 
during inflation so that the two adjacent vacua are realized randomly in each Hubble horizon.
This leads to the formation of domain walls when the Hubble parameter
becomes comparable to the axion mass, $H \sim m$.  At the same time,
the axion coherent oscillations are induced. The dynamics of axion coherent oscillations, especially its motion in the slow-roll regime, generates the baryon asymmetry in the background thermal plasma as in the usual spontaneous baryogenesis.  Let us denote the baryon asymmetry by $Y_\mathrm{osc}$. As the axion has initially large quantum fluctuations at super-horizon scales,  $Y_\mathrm{osc}$ has isocurvature fluctuations at large-scales, which is the counter part of the baryonic isocurvature fluctuations in the spontaneous baryogenesis in the slow-roll regime.
In our case, the size of the baryonic isocurvature perturbations, $\delta Y_\mathrm{osc}/Y_\mathrm{osc}$, is expected to be of order unity. This can be understood by noting that the chemical potential can be either positive or negative, depending on which vacuum the axion is rolling down to.
After the commencement of oscillations, the scalar wave dynamics between walls are random and complicated. In particular, the spatially averaged effective chemical potential is 
zero, and  no fluctuations at super-horizon scales are induced by the dynamics in the scaling regime. 
Therefore,  $Y_\mathrm{osc}$ and its fluctuations at large scales receive the main contribution from the domain wall formation when $H\sim m$. 

Finally, the domain-wall dynamics toward the scaling regime will also induce the baryon isocurvature perturbations.
For domain walls to be formed, or more precisely, for infinitely long domain walls to be formed, the probabilities to realize the two vacua must be comparable, but they do not have to be exactly equal to each other. It implies that, when domain walls are formed, the spatial volume of one of the vacua is generically larger (or smaller)
 than that of the other by (at most) a few tens of percent.
The ratio of the two volumes will quickly converge to unity as the domain-wall network approaches 
the scaling evolution. This is because the two vacua are degenerate in energy and there is no preference to one over the other once the scaling regime is reached. 
In this process toward the scaling regime, there is an overall transition from one of the vacua to the other, which similarly induces the baryon asymmetry. Let us denote the asymmetry by $Y_\mathrm{DW,form}$. As the bias of the spatial volumes is induced by the quantum fluctuations of the axion, 
$Y_\mathrm{DW,form}$ has isocurvature fluctuations at large scales.
The magnitude of $Y_\mathrm{DW,form}$ is expected to be comparable to $Y_\mathrm{osc}$, and the sign is opposite. So, there is a partial cancellation, but  in general, there is no exact cancellation. For our scenario to work, both $Y_\mathrm{osc}$ and
$Y_\mathrm{DW,form}$ must be sufficiently suppressed, since otherwise the baryonic isocurvature perturbations
and their non-Gaussianity, would be too large to be consistent with observations. 

The baryon asymmetry generated at the domain-wall formation can be suppressed as follows.  
If the lepton-number violation processes are in equilibrium between the formation and annihilation of domain walls, the initial asymmetry $Y_\mathrm{osc}$ and $Y_\mathrm{DW,form}$ can be washed out. This is the case if the reheating temperature is higher than $\sim 10^{13}$\,GeV. For lower reheating temperature, the lepton-number violating processes remain decoupled all the time. Then,  $Y_\mathrm{osc}$ and $Y_\mathrm{DW,form}$ can be suppressed if the lepton-number violating rate is much smaller than the Hubble parameter at the domain formation. 

In our numerical calculations, we have estimated $|Y_\mathrm{osc}| \sim |Y_\mathrm{DW,form}|$ by following the motion of a test domain wall which goes through a fixed position at $H = m$. Using
the test domain wall as  background classical field evolution, we have calculated the induced baryon asymmetry in the plasma by solving the Boltzmann equation. By doing so, we effectively evaluate $|Y_\mathrm{DW,form}|$ (or  $|Y_\mathrm{osc}|$) at the formation, neglecting the complicated dynamics of the scalar waves and domain-wall evolution, which do not have any preference to baryons over anti-baryons.

The current constraint on the matter isocurvature perturbation $\mathcal{S}$ 
from the Planck observation reads $\mathcal{P}_\mathcal{S} < 8.7\times 10^{-11}$~\cite{Ade:2015lrj}.
Using the fact that  baryon isocurvature perturbation is written as $\mathcal{P}_{\mathcal{S},b}^{1/2} \sim \delta \Omega_b/\Omega_m \simeq 0.15(\delta \Omega_b/\Omega_b)$, we obtain the constraint on the baryon isocurvature perturbations as $\delta\Omega_b/\Omega_b \lesssim 6 \times 10^{-5}$.
Since the baryons produced by the axion coherent oscillations or domain wall dynamics toward the scaling regime
 is $O(1)$ in the present scenario, $\Omega_{b,\mathrm{osc}}/\Omega_b \lesssim 6 \times 10^{-5}$ must be satisfied in order to avoid too large isocurvature perturbations.
Then, we obtain the constraint on the resultant baryon asymmetry induced by the coherent oscillations, 
\beq
	\frac{n_{B,\mathrm{osc}}}{s} = \frac{n_B}{s} \frac{\Omega_{b,\mathrm{osc}}}{\Omega_b} \lesssim 5 \times 10^{-15},
\eeq
and  a similar bound on the asymmetry induced by the domain wall dynamics toward the scaling regime.
This upper bound is shown by a dotted (blue) line in  Figs.~\ref{fig:contour_ab} and \ref{fig:contour_cd}.

The baryon isocurvature perturbations may be further suppressed in some particular situations.
For example, one can consider a case in which the U(1)$_{B-L}$ gauge symmetry is still unbroken
 the onset of the axion oscillation and it gets spontaneously 
 broken before the domain wall annihilation. In such a case, there is no lepton number violating operators and no baryon asymmetry is induced until the spontaneous break down of the U(1)$_{B-L}$ symmetry. 
 If the domain wall network already follows the scaling law when the U(1)$_{B-L}$ symmetry gets spontaneously 
 broken, no baryon isocurvature perturbation is generated by the coherent oscillations or domain wall dynamics. 
Interestingly, cosmic strings are  formed after the spontaneous breaking of U(1)$_{B-L}$ and they can emit a sizable amount of gravitational waves which can be within the reach of future observations \cite{Jenet:2006sv}.

\section{Discussion and Conclusions}
\label{sec:concl}

Collapsing domain walls are cosmological sources of gravitational waves~\cite{Gleiser:1998na,Hiramatsu:2010yz,Kawasaki:2011vv}.
The gravitational wave spectrum is peaked at a frequency,
\beq
	f_\mathrm{peak} \simeq 160~\mathrm{kHz}~\xi^{-1/2} \bigg( \frac{g_*}{106.75} \bigg)^{1/6} \bigg(\frac{T_X}{10^{12}~\mathrm{GeV}} \bigg),
\eeq
 corresponding to the Hubble horizon scale at the domain wall  annihilation~\cite{Hiramatsu:2013qaa}.
Here $\xi$ and $T_X$ are defined as
\beq
	\xi = \min \left( 1,~ \bigg(\frac{\Gamma_I}{H_\mathrm{ann}} \bigg)^{2/3} \right),~~ T_X = \min( T_R,~T_\mathrm{ann}).
\eeq
For successful baryogenesis, $T_X$ must be higher than $2 \times 10^{11}$\,GeV, and so, the peak frequency is at $\mathcal{O}(100)$\,kHz or higher,
which is too high to be detected by near future observations.
We note however that there have been proposed several new detection techniques with the sensitive frequency region around
MHz \cite{Nishizawa:2007tn,Goryachev:2014yra}, which may be able to probe gravitational waves produced in our scenario. 

So far, we have considered the $L$-number violating processes mediated by heavy right-handed neutrinos in the seesaw mechanism.
Other types of the baryon/lepton violating operator is also possible  and the corresponding decoupling temperature for the baryon/lepton violating 
processes could be lowered. One of the examples is the R-parity violating operator,
\beq
	W =  \frac{1}{2} \lambda_{ijk} L_i L_j \bar{E}_k
\eeq
in the supersymmetric Standard Model.
In this case, the interaction rate for the $L$-violating processes scales as $\Gamma \propto T^5$ for $T \ll m_{\tilde{\ell}}$ and $\Gamma \propto T$ for $T \gg m_{\tilde{\ell}}$, where $m_{\tilde{\ell}}$ is the  slepton mass. For instance, if we take $\lambda \sim 10^{-8}$ and  $m_{\tilde{\ell}} \gtrsim 10^{9}$ GeV, the $L$-violating
process marginally reaches equilibrium and soon decouples at $T_\mathrm{dec} \sim 10^9$\,GeV.
Since the maximal possible value of lepton asymmetry is roughly given by $n_L/s \sim 0.1 T_\mathrm{dec}/M_P$ from the 
first equality in (\ref{eq:nLs_max}), successful baryogenesis is possible with $T_\mathrm{ann} \sim 10^9$ GeV.
In this case, the peak frequency of the gravitational waves from the domain wall annihilation can be within the sensitivity range of the ground-based 
detector such as advanced-LIGO \cite{Abramovici:1992ah} and KAGRA \cite{Somiya:2011np,Aso:2013eba}.
For instance, if we take $T_R \sim m \sim 10^9$~GeV and $f \sim 10^{13}$~GeV, domain walls dominate the Universe at the annihilation 
and the peak frequency falls in the sensitivity range of these experiments. A naive order-of-magnitude estimate suggests, however,  that the signal strength is
a few orders of magnitude smaller than the predicted sensitivity, and either some deviation from the scaling regime or further improvement of the sensitivity
would be necessary to directly probe such signals.

In this paper we have proposed a baryogenesis scenario using axion domain walls. Axion domain walls are produced if the axion
acquires sufficiently large quantum fluctuations during inflation or if it initially stays sufficiently close to the local maximum.
While no net baryon asymmetry is produced in the scaling regime, 
collapsing axion domain walls produce a large enough baryon asymmetry to explain the observed value. This is because the energy bias
between the two vacua, and therefore between baryons and anti-baryons, becomes relevant only when domain walls annihilate.
In particular,  baryon isocurvature perturbations can be significantly suppressed in our scenario, either because the asymmetry produced
by the initial field configurations is washed out by the $L$-number violating interactions in equilibrium, or because the $L$-number violating
interaction is simply suppressed at the domain wall formation. In some parameter region,  baryon isocurvature perturbations
and their non-Gaussianity are suppressed, but non-negligible, which may be detected by future observations.
Our scenario works together with high-scale inflation which predicts a large tensor-to-scalar ratio within the reach of future B-mode observations.
The required relatively high reheating temperature can be realized in high-scale inflation more easily. This should be contrasted to other spontaneous baryogenesis
scenarios in which the inflation scale is severely constrained by the isocurvature perturbations. 
Although we have focused on the axion domain wall throughout this paper, our analysis can also be straightforwardly applied to a wide class of domain walls 
such as the Standard Model Higgs domain wall \cite{Kitajima:2015nla,Kusenko:2014lra}.

\section*{Acknowledgment}
This work was supported by  JSPS Grant-in-Aid for
Young Scientists (B) (No.24740135 [FT]), 
Scientific Research (A) (No.26247042 [FT]), Scientific Research (B) (No.26287039 [FT]), and
 the Grant-in-Aid for Scientific Research on Innovative Areas (No.23104008 [NK, FT]). 
This work was also supported by World Premier International Center Initiative (WPI Program), 
MEXT, Japan [FT].

\end{document}